\documentclass[prl, twocolumn, showpacs, preprintnumbers,amsmath,amssymb,superscriptaddress]{revtex4-1}
\usepackage{graphicx}
\usepackage{dcolumn}
\usepackage{bm}
\usepackage{subfigure}
\usepackage{multirow}
\usepackage{amsmath}
\usepackage{color,soul}
\usepackage[font=footnotesize, justification=raggedright,singlelinecheck=false]{caption}
\begin{document}

\title{Non-Equilibrium Strongly Hyperuniform Fluids of \\\  Circle Active  Particles  with  { Large Local Density Fluctuations}}

\author{Qun-Li Lei}
\affiliation{School of Chemical and Biomedical Engineering, Nanyang Technological University, 62 Nanyang Drive, 637459, Singapore}

\author{Massimo Pica Ciamarra}
\email{massimo@ntu.edu.sg}
\affiliation{Division of Physics and Applied Physics, School of Physical and Mathematical Sciences, \\ Nanyang Technological University, 21 Nanyang Link, 637371, Singapore
}

\author{Ran Ni}
\email{r.ni@ntu.edu.sg}
\affiliation{School of Chemical and Biomedical Engineering, Nanyang Technological University, 62 Nanyang Drive, 637459, Singapore}

\begin{abstract}
Disordered hyperuniform structures are an exotic state of matter having vanishing long-wavelength density fluctuations similar to perfect crystals but without long-range order. Although its importance in materials science has been brought to the fore in past decades, the rational design of experimentally realizable disordered strongly hyperuniform microstructures remains challenging. Here we find a new type of non-equilibrium fluid with strong hyperuniformity in two-dimensional systems of chiral active particles, where particles perform independent circular motions of the radius R with the same handedness. This new hyperuniform fluid features a special length scale, i.e., the diameter of the circular trajectory of particles, below which large density fluctuations are observed. By developing a dynamic mean-field theory, we show that the large local density fluctuations can be explained as a motility-induced microphase separation, while the Fickian diffusion at large length scales and local center-of-mass-conserved noises are responsible for the global hyperuniformity.
\end{abstract}

\maketitle
\section{Teaser}
{Dynamic hierarchical structures with strong hyperuniformity are found in chiral active matter systems.}

\section{Introduction}
Perfectly ordered structures, like crystals or quasi-crystals {at zero temperature}, are usually associated with some discrete symmetries and exhibit long-range correlations, leading to the structure factor of the system $S(q \rightarrow 0) = 0$~\cite{torquato2018PR}. Similarly, the local density variance in these structures $\langle \delta \rho^2 \rangle$ scales with the window size of observation $L$ as $\langle \delta \rho^2 \rangle \sim L^{-\lambda}$ with $\lambda = d+1$, where $d$ is the dimensionality of the system. In contrast, in normal disordered structures, e.g., conventional gases, liquids and amorphous solids {and even thermalized crytals}, the long wavelength density fluctuation makes $S(q\rightarrow 0) = const. > 0$  and $\lambda = d$.  Recently, the concept of hyperuniformity was introduced to study the state of matter~\cite{torquato2003local}. A structure is hyperuniform when it has vanishing long wavelength density fluctuations, i.e, $S(q \rightarrow 0) \sim q^{\alpha} \rightarrow 0$ with $\alpha > 0$ and $\langle \delta \rho^2 \rangle \sim L^{-\lambda}$ with $d < \lambda \le d+1$~\cite{torquato2003local}.  
It has been found in the past two decades that, besides the ordered hyperuniform structures, i.e., perfect crystals and quasi-crystals, a number of \emph{disordered} structures are also hyperuniform, including the maximally random jammed packing~\cite{donevprl2005}, avian photoreceptor patterns~\cite{jiaopre2014} and some non-equilibrium  systems~\cite{hexner2015,tjhung2015,
hexner2017enhanced,weijs2015emergent,
wang2017hyperuniformity,hexner2017noise}. 

Disordered hyperuniform structures have received an increasing amount of scientific attention, as some strongly hyperuniform disordered structures {with $\lambda = d+1$} exhibit similar properties as crystals with even better performance, e.g., large isotropic photonic band gaps insensitive to defects~\cite{florescu2009designer} can be opened at low dielectric contrast~\cite{man2013photonic}. {Although ideal hyperuniform structures similar to perfect crystals are unavoidably affected by thermal excitation, it still shows promise in designing robust disordered materials with novel functionalities~~\cite{kim2018effect,
edagawa2008photonic,xie2013silicon,leseur2016high,torquato2018PR}.}
By far, various protocols were developed to design particle interactions to form disordered hyperuniform ground states in classical many particle systems. However,  the resulting interactions normally have delicate long-range or multi-body terms~\cite{torquato2018PR}, making the experimental realization highly challenging.  An alternative approach is to use driven systems, e.g., self-organized colloidal suspension under periodic shearing~\cite{corte2008random}, to form non-equilibrium dynamic hyperuniform states {which can effectively avoid the dynamic trapping and in principle has a higher resistance to thermal perturbations~\cite{hexner2017enhanced}.} Currently, experimentalists only succeeded in generating weakly hyperuniform structures with $\lambda \simeq d+0.5$~\cite{weijs2015emergent} using this method. Nevertheless, this is certainly a  direction that deserves further investigation~\cite{hexner2017noise}, as some self-driven systems, or active matter systems, have produced a number of strikingly surprising emergent phenomena never found in corresponding equilibrium systems~\cite{marchetti2013,ramaswamy2003active,
fily2012athermal,speck2014effective,niprl2015}. { However, in conventional active matter systems, i.e., active nematic systems and active Brownian particles systems, giant number fluctuations characterized by $\lambda < d$ ~\cite{ramaswamy2003active} and motility-induced phase separation (MIPS) with $\lambda \simeq  0$ ~\cite{fily2012athermal,lowen2013} are usually observed. These large density fluctuations seemingly prohibit the formation of hyperuniform structures in  active matter systems}.

Recently, chiral active matter whose motion is chiral-symmetry broken, e.g., active particles/swimmers with circular motion {in 2D}~\cite{lauga2006,
lowen2016,han2017effective,zhou2017twists,wioland2016} or active spinner fluids~\cite{scholz2017velocity} has attracted considerable attention. Both experiments and simulations have shown many interesting collective phenomena in these systems~\cite{lowen2016,ma2017driving,liebchen2017collective,chen2017weak,souslov2017topological}.  In this work, by using computer simulations combined with analytic
theories, we study a 2D system of circle active particles, which perform independent circular motion with the same handedness and random circling phases. We show that in the limit of strong driving  {or zero thermal noise}, with increasing the density of particle or the radius of circular motion $R$, the system undergoes an absorbing-active transition forming a non-equilibrium strongly hyperuniform fluid phase with density variance $\langle \delta \rho^2 \rangle \sim L^{-3}~ (L\rightarrow \infty)$ as the same as {perfect} crystals. Further increasing the density or $R$ triggers {the  formation of dynamic clusters,} which results in large local density fluctuations. These fluctuations are `confined' within the length scale of $R$, while the strong hyperuniformity persists at large length scales. This surprising coexistence of {large local density fluctuations} and the global hyperuniformity is explained by dynamic mean-field theories at different length scales.

\begin{figure*}[htbp] 
\centering
\begin{tabular}{c}
	\resizebox{150mm}{!}{\includegraphics[trim=0.0in 0.0in 0.0in 0.0in]{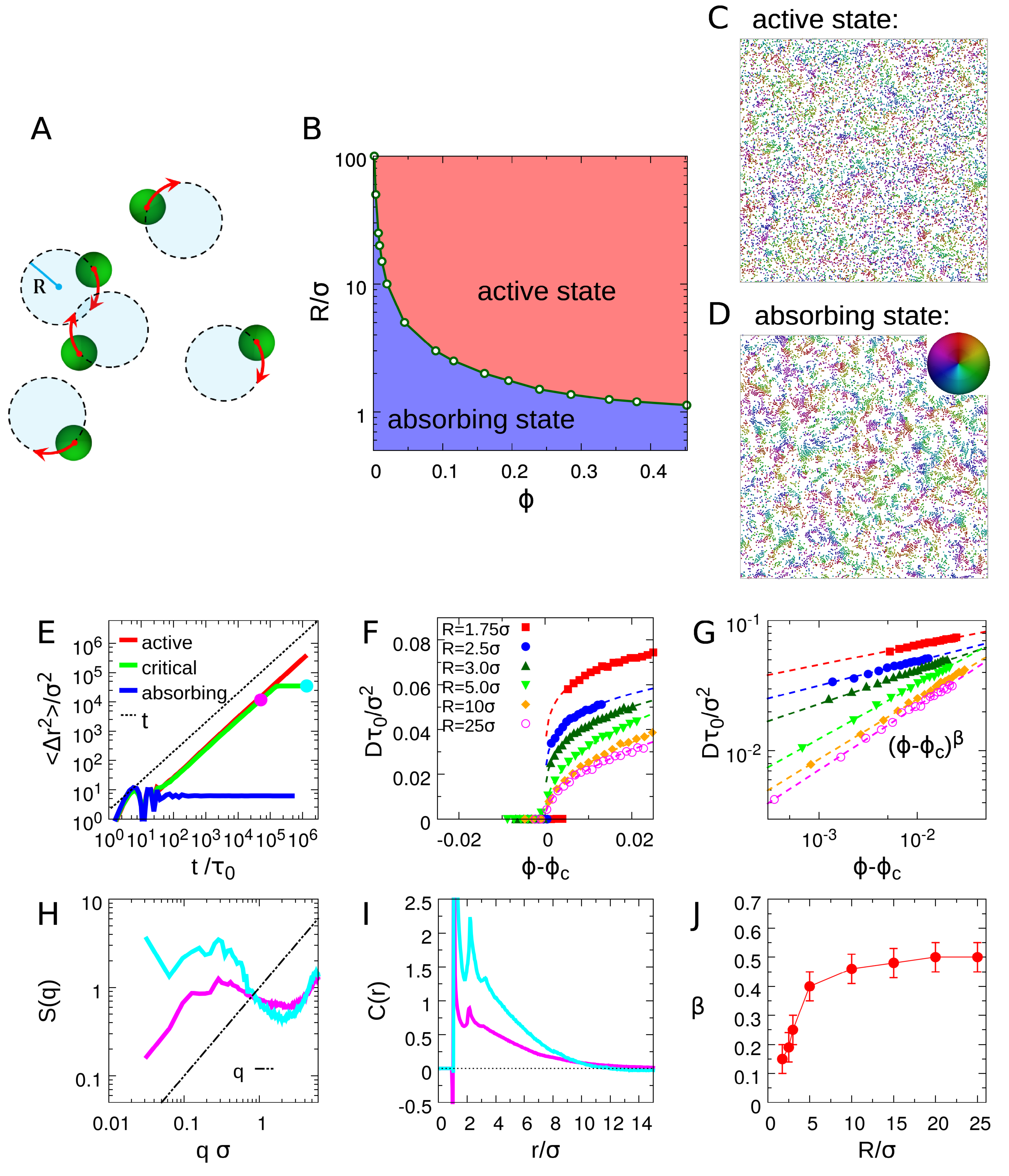} }
\end{tabular}
\caption{{\bf Absorbing-active transition.} (\textbf{A}): Schematic of the 2D system of circle active particles. (\textbf{B}): Dynamic phase diagram in the representation of packing fraction $\phi$ and circle radius $R$. (\textbf{C,~D}): Typical snapshots of the active and adsorbing states near the critical point with $\phi=0.20$ with $R=1.75\sigma$, where the color indicates the self-propulsion orientation of each particle. These two states are marked as magenta and cyan solid symbols respectively in \textbf{E}. The structure factor $S(q)$ and the orientation correlation function $C(r)$ of these two states are shown in (\textbf{H,~I}). { (\textbf{E}): MSD as functions of time for system with $R=1.75
\sigma$ started from random configurations. Red line: active state ($\phi=0.22$); blue line absorbing state ($\phi=0.01$); green line: system near the critical point ($\phi=0.20$) in which the system ultimately falls into the absorbing state after a long simulation time. (\textbf{F,~G}): Diffusion constant as functions of $\phi$ near the critical point $\phi_c$ for systems with different $R$. The dashed lines are the fitting of power law $(\phi-\phi_c)^\beta$. (\textbf{J}): The measured critical exponent $\beta$ as a function of $R$. For all the calculations, $N = 10,000$ and $T_{R}=0$.}}
\label{Fig_schematic}
\end{figure*} 

\section{Results}

\subparagraph{Model}
As illustrated in Fig.~\ref{Fig_schematic}A, we consider a 2D suspension of $N$  active colloidal particles with diameter $\sigma$. Each particle experiences an in-plane force $F^p$ with random initial orientation as well as a constant torque $\mathbf{\Omega}$ perpendicular to the plane, which drive the particles to perform circular motion with the same handiness~\cite{ma2017driving,liebchen2017collective}.  The dynamics of particle $i$ at finite temperature $T$ is governed by the over-damped Langevin equations~\cite{ma2017driving,han2017effective}
\begin{eqnarray}
\dot{\mathbf{r}}_i(t) &=& \gamma_t^{-1} \left[-{\nabla}_i U(t) + F^p \textbf{e}_i(t)\right] + \sqrt{2 k_B T/\gamma_t}~{\bm \xi}^t_i(t),~~~ \label{eq1}
\\
\dot{\textbf{e}}_i(t) &=&  \left[  \gamma^{-1}_r {\bm \Omega} + \sqrt{2 k_B T/\gamma_r}~{\bm \xi}^r_i(t) \right] \times {\textbf{e}}_i(t),\label{eq2}
\end{eqnarray}
where $\mathbf{r}_i$ and $\mathbf{e}_i$ are the position of particle $i$ and its self-propulsion orientation, respectively. $\gamma_{t/r}$ is the translational/rotational friction coefficient. For simplicity, we set $\gamma_{t}= \gamma_{r}/\sigma^2$.  ${\bm \xi}^t_i(t)$ and ${\bm \xi}^r_i(t)$ represent Gaussian noises with zero mean and unit variance. We use Weeks-Chandler-Andersen (WCA) potential to mimic the excluded volume interaction between colloidal particles $U(t)$ (see the Method). The packing fraction of the system is defined as $\phi= \rho\sigma^2\pi/4$ with $\rho $ the particle density. The self-propulsion speed of particle is $v_0=\gamma_t^{-1} F^p$. The {reduced noise strength in the system is defined as $T_{R} = {k_BT}/({F^p\sigma})$} which measures the strength of thermal noise compared with the self-propulsion.  In the {zero noise limit, i.e., $T_{R} = 0$,} {isolated active particles} perform circular motions with fixed radius $R=F^p\sigma^2/\Omega$ and period $\Gamma=2\pi\gamma_r/\Omega$. {This athermal noise-free situation is the major focus of this work and the effect of thermal noise is discussed later.}

\subparagraph{Absorbing-active transition}
We first simulate systems with $N=10,000$ and $T_{R}=0$. Under low packing fraction $\phi$ and small $R$ condition, we find that the system falls into an absorbing or arrested state, in which each particle performs independent circular motion without collisions, and the mean squared displacement (MSD) $\langle \Delta r^2 \rangle$ of particles develops a plateau at long time (see the blue line in  Fig.~\ref{Fig_schematic}E). With increasing $\phi$ or $R$, the collisions between particles become more frequent, making the system unable to find an non-interacting state. Thus, the system remains at an active diffusive state with MSD $\langle \Delta r^2 \rangle \sim 4 D t$ ($t \rightarrow \infty$) (red line in Fig.~\ref{Fig_schematic}E). Here $D$ is the long-time diffusion constant.   {The phase behaviors of the system are summarized in the phase digram Fig.~\ref{Fig_schematic}B. In the following, we  focus on the absorbing-active transition close to the boundary of two phases.}

 {In Fig.~\ref{Fig_schematic}F, we plot $D$ as a function of $\phi-\phi_c$ for different $R$. Here $\phi_c$ is obtained by fitting with the critical power law $D \sim (\phi-\phi_c)^{\beta}$, which determines the phase boundary in Fig.~\ref{Fig_schematic}B.  One can observe a sharp transition from the absorbing state ($D(\phi) =0$) to the active state ($D(\phi) > 0$) when increasing $\phi$ at a small $R=1.75\sigma$. The transition becomes relatively smoother as the $R$ increases.  In Fig.~\ref{Fig_schematic}G, we show the log-log plot of $D$ as a function of $\phi-\phi_c$. The obtained slope $\beta$ is given in  Fig.~\ref{Fig_schematic}J. We find the critical exponent $\beta$ is about $0.15$ for $R=1.75\sigma$, which is substantially smaller than the values in classical absorbing transitions, i.e., $\beta=0.58$ for directed percolation and $\beta =0.64$ for conserved directed percolation~(Manna type)~\cite{hinrichsen2000non,tjhung2015}. Such a small critical exponent is independent of system size (see Supplementary Fig. S1). With increase $R$, we find that $\beta$ increases to around 0.5  for $R>10\sigma$.} {Similar increase of critical exponent with increasing the interaction range (in our case, $R$) has been reported~\cite{lubeck2003}.} In our system, $\phi_c$ would decrease to zero with increasing $R$. Hence, {$\beta$} at large $R$ can not be directly obtained in our system due to the divergence of simulation time needed at the dilute limit.

 {  To understand the physics behind the absorbing-active transition at small $R$, we choose a packing fraction $\phi=0.20$ close to the critical point ($\phi_c=0.195$)  for system with $R=1.75\sigma$. The MSD for the system started from random configuration is shown by the green line in Fig.~\ref{Fig_schematic}E, in which one can see that the system ultimately falls into the absorbing state after staying at the active state for a long time. In Fig.~\ref{Fig_schematic}C,D, we show typical snapshots for the active state and absorbing state before and after the absorbing transition as indicated by the magenta and cyan symbols in Fig.~\ref{Fig_schematic}E. Movies for these two states can be found in Supplementary movie 1-2.} One can notice a marked structural difference between these two states. In the absorbing state, particles with similar orientation form finite {synchronized} clusters, while the active state is more homogeneous without  {noticeable} spatial heterogeneity. This structural difference is also reflected in the structure factor $S(q)$ and orientation correlation function $C(r)= \langle \sum_{i \ne j}  \mathbf{e}_i\cdot \mathbf{e}_j ~ \delta(r_{ij}-r) \rangle/\rho N$ as shown by Fig.~\ref{Fig_schematic}H,I, respectively. Compared with the active state, $S(q)$ for the absorbing state develops a pronounced peak at $q \sigma \simeq 0.2$ and the corresponding $C(r)$ also shows a {stronger} orientation correlation. {The {synchronized} clusters formation in our circle active particle system with isotropic circling-phase distribution shares a similar mechanism with the phase separation observed in an experimental bimodal phase-distributed system~\cite{han2017effective}(see also Supplementary Fig.~S4). Both are a result of an crowding-induced attraction between particles with the same circling phase. We find that this distinct structural transformation during the absorbing transition is absent in the conventional absorbing transition~\cite{corte2008random,hexner2015,tjhung2015,hinrichsen2000non}, suggesting that the small critical exponent measured in our system has a structural origin. With increase the $R$, the structural difference between two phases becomes weaker (see Supplementary Fig. S2) which occurs simultaneously with the increase of $\beta$.  Further studies combining with finite size analysis are necessary for determining whether the absorbing-active transition at small $R$ is first-order.}


\begin{figure*}[htbp] 
\centering
\begin{tabular}{c}
	\resizebox{180mm}{!}{\includegraphics[trim=0.0in 0.0in 0.0in 0.0in]{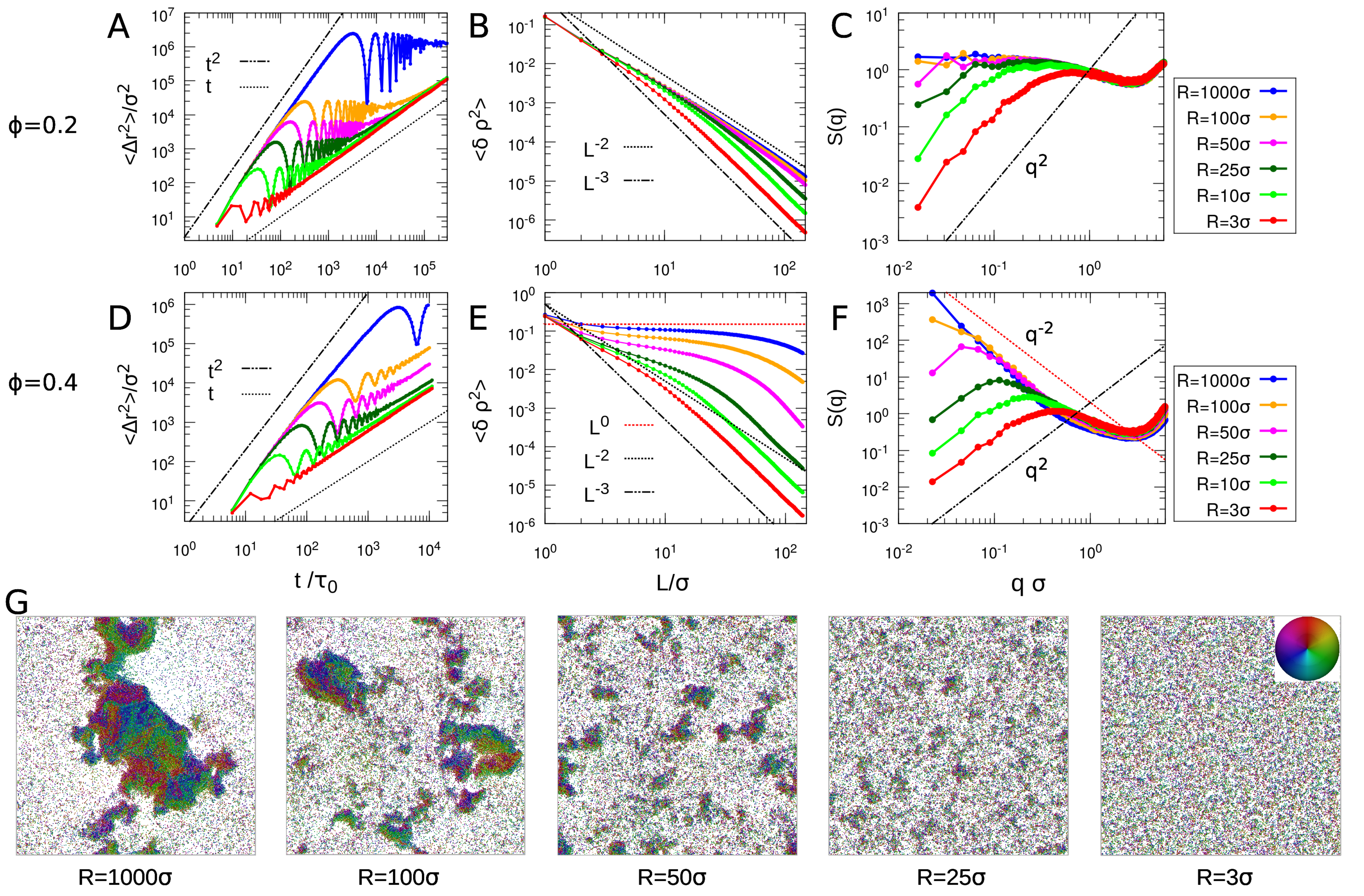} }
\end{tabular}
\caption{{\bf Dynamic hyperuniform state.}
(\textbf{A,D}): MSD as functions of $t$ for various $R$.
(\textbf{B,E}): Density variances $\langle \delta \rho^2 \rangle$ as functions of window size $L$ for various $R$. The $L^{-3}$ asymptotic line indicates the hyperuniform scaling which is as the same as in perfect crystals. The $L^{-2}$ scaling is for normal fluids, while $L^{0}$ is for { clustering or phase separation} induced large density fluctuations.
(\textbf{C,F}): Structure factor $S(q)$ for various $R$. The $q^2$ asymptotic line indicates the hyperuniform scaling, while the $q^{-2}$ line represents { clustering or phase separation} induced large density fluctuations.
(\textbf{G}): Typical snapshots for systems at $\phi = 0.4$ with various $R$. For all the calculations, $N = 40,000$ and $T_{R}=0$.}
\label{Fig_hyperuniform}
\end{figure*}

\begin{figure*}[htbp] 
\centering
\begin{tabular}{c}
	\resizebox{130mm}{!}{\includegraphics[trim=0.0in 0.0in 0.0in 0.0in]{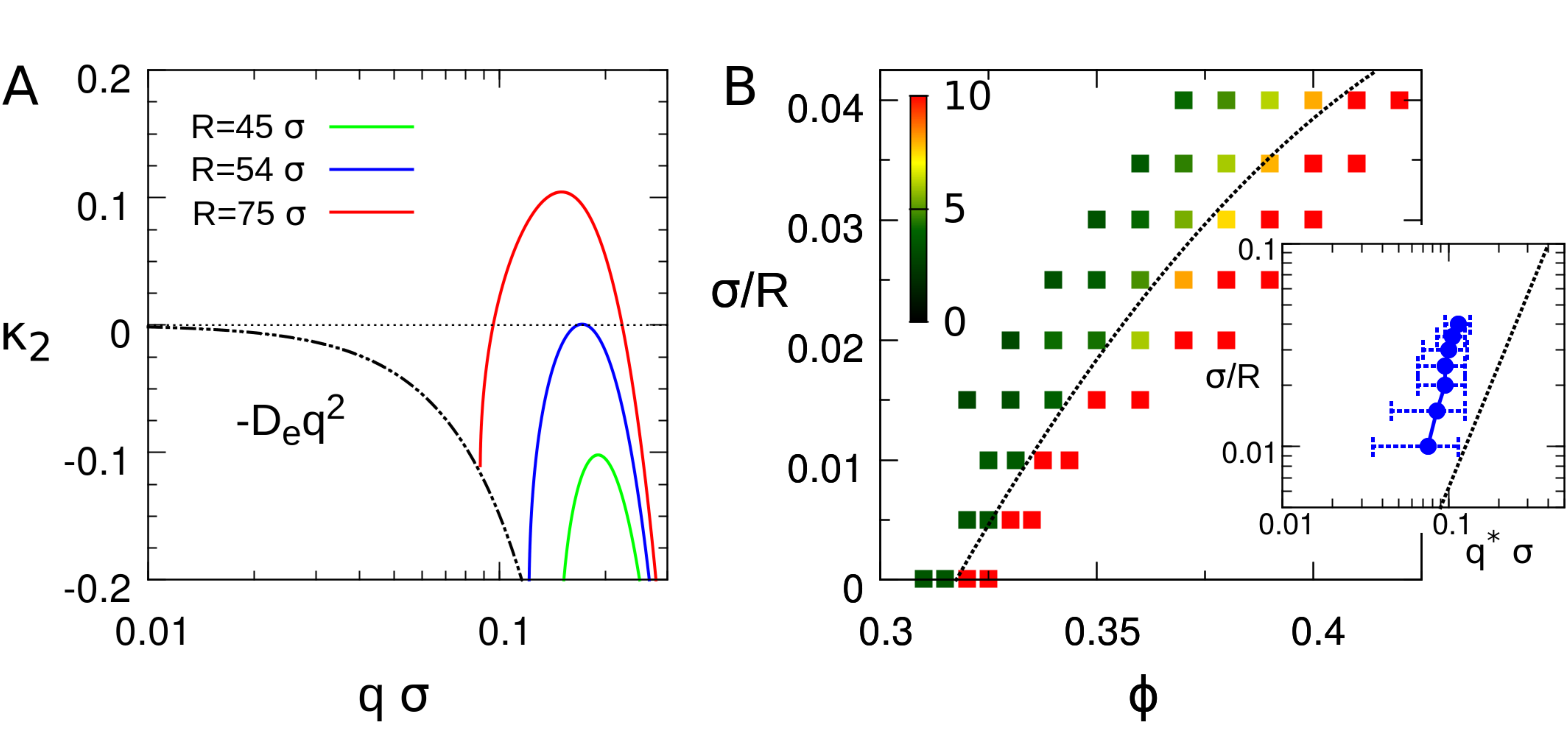} }
\end{tabular}
\caption{{\bf {Dynamic microphase separation}.} (\textbf{A}): Growing rate $\kappa_2$ as functions of $q$ for systems at $\phi = 0.35$ with various $R$ obtained from the dynamic mean-field theory.
(\textbf{B}): Measured heights of the first peak in $S(q)$ for systems with different combinations of $R$ and $\phi$ in computer simulations indicated by the color of the symbols. The dotted line is the fitting using Eq.~\ref{critital_phi} for the phase boundary. Inset: the measured position of the first peak in $S(q)$ in computer simulations (symbols) and the theoretical prediction (dotted line) based on the fitting in \textbf{B}.}
\label{Fig_dispersion}
\end{figure*}

\subparagraph{Hyperuniformity and {large local density fluctuations}}
As shown in Fig.~\ref{Fig_schematic}H, for the active state near the critical point in system with $R=1.75\sigma$, the structure factor of the system exhibits hyperuniform scaling $S(q)\sim q$ ($q\rightarrow 0$). In the random organization model aiming to mimic the colloidal suspension under periodic shearing, a similar hyperuniform  scaling was observed near the critical point~\cite{hexner2015,tjhung2015}. {However, for relative large $R$, the critical $q$ scaling shifts to a faster $q^2$ scaling (see Supplementary Fig. S2).} To explore this further, we simulate a system of $N=40,000$ circle active particles. In Fig.~\ref{Fig_hyperuniform}A, we first plot the MSD for active state systems with different $R$ at $\phi=0.2$. We find the diffusivity in the system rises slightly with increasing $R$. By further checking the scalings of the density variance $\langle \delta \rho^2 \rangle$ and $S(q)$ for different $R$ (see Fig.~\ref{Fig_hyperuniform}B,C), we observe a strong hyperuniformity in the systems {with $R\leq 25 \sigma $}, as indicated by the asymptotic behaviours: $\langle \delta \rho^2 \rangle \sim L^{-3}$ ($L \rightarrow \infty$) and $S(q) \sim q^2$ ($q \rightarrow 0$). From $\langle \delta \rho^2 \rangle$  one can identify a $R$ dependent length scale $L_{HU}$, above which the system becomes hyperuniform, while below which the system behaves like normal fluids, i.e., $ \langle \delta \rho^2 \rangle \sim L^{-2} $ (Fig.~\ref{Fig_hyperuniform}B).  In $S(q)$, a similar  threshold $q_{HU}$ can be found, which features the end of the hyperuniform scaling.  With increasing $R$, $L_{HU}$ increases and $q_{HU}$ decreases { until $R\geq 50\sigma$ where the hyperuniform scaling }{ becomes less apparent due to the system finite size effect as discussed below.} This implies that $R$ controls the length scales at which the system exhibits hyperuniformity.

Figures~\ref{Fig_hyperuniform}D-F show the result of an analogous investigation for higher density systems with $\phi = 0.4$. Compared with low density systems, pronounced enhancement of the diffusivity with increasing $R$ is observed (Fig.~\ref{Fig_hyperuniform}D). {  The high density systems also show clear hyperuniform scaling at large length scales for $R\leq 50\sigma$} and the threshold $L_{HU}$ (or $q_{HU}$) increases (or decreases) with increasing $R$ (Fig.~\ref{Fig_hyperuniform}E,F). However, at $\phi=0.4$, we find  $\langle \delta \rho^2 \rangle \sim L^{-\lambda}$ with $\lambda < 2$ when $L \ll L_{HU}$ and $\lambda$ decreases with larger $R$. This implies that at length scales $L \ll L_{HU}$, the system exhibits {large density fluctuations}, whose strength and length scale  are both controlled by $R$. This large fluctuation can also be identified by the scaling $S(q) \sim q^{-2}$ for $q > q_{HU}$ as shown in Fig.~\ref{Fig_hyperuniform}F, which was reported as a signature of critical instability of active particle system undergoing MIPS~\cite{fily2012athermal}. However, in most of our systems except $R \geq 100\sigma$, $S(q) \sim q^{-2}$ does not diverge at $q_{HU}\simeq 0$ as in MIPS~\cite{fily2012athermal}, but stops at a finite $q_{HU}$. This suggests that the {large local density fluctuations} observed in our system are because of  {clustering or microphase formation}. The crossover of two different scalings, i.e., {large density fluctuations} and hyperuniformity, creates  a peak in $S(q)$ at $q_{HU}$, whose height increases with larger $R$. {  Actually for $R \geq 100\sigma$, we speculate that in much larger systems, one can still observe the peak at finite $q_{HU}$ for $\phi=0.4$, as well as the hyperuniform scaling for both $\phi=0.2$ and 0.4, as suggested by later theoretical analyses. In Supplimentary Fig. S3, $S(q)$ is shown for a larger system ($N=102,400$) at $\phi=0.4$. 
The result agrees with our speculation.} Typical snapshots of the system at $\phi=0.4$ with various $R$ are shown in Fig.~\ref{Fig_hyperuniform}G (also Supplementary movie 3-7 and Fig S3), and one can see many finite-size clusters  disappearing and reforming in the system. The average size of these dynamic clusters increases with increasing $R$, and at $R=1000\sigma$, because of the finite size effect, the clusters percolate the simulation box. These findings are intriguing, as hyperuniformity and large density fluctuations induced by {dynamic cluster formation} are two seemingly opposite phenomena, which coexist here in the same system at different length scales. In the following, we formulate dynamic mean-field theories to understand this new dynamic hyperuniform fluid with {large local density fluctuations}.

\subparagraph{Dynamic mean-field theory}
Starting from the  $N$-body Smoluchowski equation for active Brownian particles~\cite{fily2012athermal,speck2014effective}, one can prove  (Supplementary Information S1) that in a homogeneous circle active particles system with vanishing orientation order parameter $\mathbf{Q}= \langle \mathbf{e}  \mathbf{e}^T - \frac{1}{2}{\bm 1} \rangle$, the time-dependent local density field $\rho(\mathbf{r}, t) $ and the local polarization field $\mathbf{p}(\mathbf{r}, t)$ satisfy
\begin{eqnarray}
\partial_t \rho  &=&  -\nabla \cdot \left[ v_e(\rho) \mathbf{p}  - D_e  \nabla  \rho  \right],  \label{eq_rho_dym0}\\
\partial_t \mathbf{p} &=&   - \frac{1}{2}\nabla [v_e(\rho)\rho]  + D_e \nabla^2 \mathbf{p}  +  \mathbf{\Omega}_r \times \mathbf{p}, \label{eq_p_dym0}
\end{eqnarray}
where $v_e(\rho)=v_0+ \zeta \rho $ is a density-dependent effective velocity of particles with a negative $\zeta$ reflecting the  motility-induced `self-trapping' effect. $D_e$  is the effective diffusion constant originated from the `evasive' motion of particles due to the collisions with neighbouring particles~\cite{speck2014effective}. $\mathbf{\Omega}_r= \gamma_{r}^{-1} \mathbf{\Omega}$ is the reduced torque.  The isotropic homogeneous state, i.e., $[\rho(\mathbf{r},t)= \overline{\rho},~ \mathbf{p}(\mathbf{r},t)= 0]$, is the solution to Eqs.~\ref{eq_rho_dym0} and \ref{eq_p_dym0}. By making a weak perturbation around this state, i.e.,
$[\rho(\mathbf{r},t)= \overline{\rho}+\delta\rho(\mathbf{r},t),~\mathbf{p}(\mathbf{r},t)=\delta \mathbf{p}(\mathbf{r},t) ]$, we obtain two linearized equations in the Fourier space with the first-order approximation
\begin{eqnarray}
\left ( i\omega +D_eq^2 \right) \delta \rho_{\mathbf{q},\omega} 
&=& -i v_e \mathbf{q}\cdot \mathbf{p}_{\mathbf{q},\omega}, \label{eq_rho_dym1}
\\
\left( i \omega + D_e q^2\right) \mathbf{p}_{\mathbf{q},\omega} 
&=& \mathbf{\Omega}_r \times \mathbf{p}_{\mathbf{q},\omega} - i {\rm w} \mathbf{q} ~\delta \rho_{\mathbf{q},\omega},  \label{eq_p_dym1}
\end{eqnarray}
where $ [\delta \rho_{\mathbf{q},\omega}, ~ \mathbf{p}_{\mathbf{q},\omega}]=\int d \mathbf{r}\cdot e^{-i\mathbf{q} \cdot \mathbf{r}} \int d t ~ e^{-i \omega {t}}[\delta \rho ,~ \mathbf{p} ]$ and ${\rm w}= v_0/2 +\zeta \overline{\rho}$ is the parameter indicating the strength of self-trapping effect.
By solving Eqs.~\ref{eq_rho_dym1} and \ref{eq_p_dym1}, one obtains the dispersion relationship which includes a diffusive mode $\omega_{0} = i D_e q^2$ and two non-diffusive modes
\begin{eqnarray} \label{eq_dominant_mode}
\omega_{1,2} &=& i D_e q^2 \pm \sqrt{v_e {\rm w}  q^2  +\Omega_r^2}.
\end{eqnarray}
The growth rate of the mode is $\kappa = {\rm Re}(i \omega)$, whose sign determines whether the perturbation $\delta \rho \sim e^{i\omega t + i {\mathbf q} \cdot {\mathbf r}}$ grows or decays. One can prove that the mode 1 always decays, while the mode 2 may grow for $\rm w<0$ with
\begin{eqnarray}
\kappa_2 & = & \left\{
\begin{array}{ll}
-D_e q^2 ~~~~~~~~~~~~~~~~~~~~~~~~~~~~~~~~  q < \frac{v_0}{ \sqrt{-v_e {\rm w}} }R^{-1}~~~~~
       \\
-D_e q^2 + \sqrt{-v_e {\rm w}  q^2  -\Omega_r^2} ~~~~~~~~~    q >\frac{v_0}{ \sqrt{-v_e {\rm w}} }R^{-1}  ~~~~~
\end{array}
\right.   \label{eq_grow_rate}
\end{eqnarray}
In Fig.~\ref{Fig_dispersion}A, we plot three typical situations of $\kappa_2$ as functions of $q$ at high density ($\rm w < 0$) by varying $R$ (or $\Omega_r$). One can see that for finite $R$, $\kappa_2$ decreases from zero following the typical diffusive mode $-D_eq^2$ from $q=0$ to $\frac{v_0}{ \sqrt{-v_e {\rm w}} }R^{-1}$, above which $\kappa_2$ starts to increase and  has the chance to be positive at finite $q>0$. This implies that the homogeneous system becomes unstable as a result of the growing fluctuation of finite wavelength.  {In the mean-field picture, this typically signatures that the system undergoes a microphase separation.} This scenario is different from the complete phase separation in which instability starts from the infinite wavelength at $q=0$~\cite{fily2012athermal,speck2014effective}.

 The instability point of the system is defined as $\kappa^{max}_2(q^*)=0$ for $q^*>0$, from which we can obtain the relationship between the critical packing fraction $\phi^*$ and $q^*$ (see Supplementary Information S2)
\begin{eqnarray}
\left(\frac{\phi^*}{\phi_c}-1\right)\left(2-\frac{\phi^*}{\phi_c}\right) = \frac{8D_e}{v_0 R}, \label{critital_phi} \\
 q^* = \sqrt{\frac{v_0}{R D_e} } ,\label{critital_R}
\end{eqnarray}
where $\phi_c\simeq 0.32$ is the critical packing fraction for $ R= \infty$ { at which the simulation shows complete MIPS ($q^* =0$). This suggests that  MIPS can be seen as a limit of the microphase separation in our system when $ R \rightarrow \infty$} { at the mean-field level.}  For this special case, when $\phi$ approaches $\phi_c$, $S(0)$ is supposed to jump from a finite value to $\infty$, which marks the `spinodal' of MIPS. For systems with finite $R$, the first peak of $S(q)$, which is located around $q^*$, is also expected to jump upto a higher value when the system crosses the {``microphase separation point"}.  In Fig.~\ref{Fig_dispersion}B, we plot the measured heights of the first peak in $S(q)$ for various combinations of $R$ and $\phi$ in computer simulations. We find when $R > 50\sigma$, the height of the first peak in $S(q)$ jumps from below 5 to above 10 with increasing $\phi$ over $\phi^*$. The transition becomes smoother for systems of smaller $R$.  Therefore, we choose $S(q) = 7$  as the threshold to fit the phase boundary using Eq.~\ref{critital_phi} (the dotted line in Fig.~\ref{Fig_dispersion}B), which leads to $D_e \simeq 0.62 v_0 \sigma$. In the inset of Fig.~\ref{Fig_dispersion}B, we plot the location of $q^*$ measured in simulations as well as the prediction of Eq.~\ref{critital_R} with the same $D_e$. The simulation results agree almost quantitatively with the theoretical prediction. {For small $R$ cases ($R< 50\sigma$), there is a noticeable inconsistency, suggesting the possible breakdown of the mean-field theory. In these cases, the size of clusters in the system is comparable with the particle size. Therefore, `microphase separation' is no longer a proper concept to describe what happens in the system}. {In fact, even for large $R$ cases, which seem consistent with the mean-field microphase separation scenario, it still remains open whether there exist true sharp microphase separation points for finite $R$ or the sharpness of transition diverges at $R\rightarrow\infty$.}

To summarize, this theoretical analysis rationalizes the instability in the circle active particle system as a result of motility-induced self-trapping effect (negative $\rm w$ in Eqs.~\ref{eq_dominant_mode},~\ref{eq_grow_rate}). More importantly, the existence of non-zero torque $\Omega_r$ restricts the growth of density fluctuations within a finite wavelength proportional to the circle radius $R$, which is the underlying reason for {the dynamic clusters formation} and the resulting large local density fluctuations observed in our simulations.

\begin{figure*}[htbp] 
\centering
\begin{tabular}{c}
	\resizebox{110mm}{!}{\includegraphics[trim=0.0in 0.0in 0.0in 0.0in]{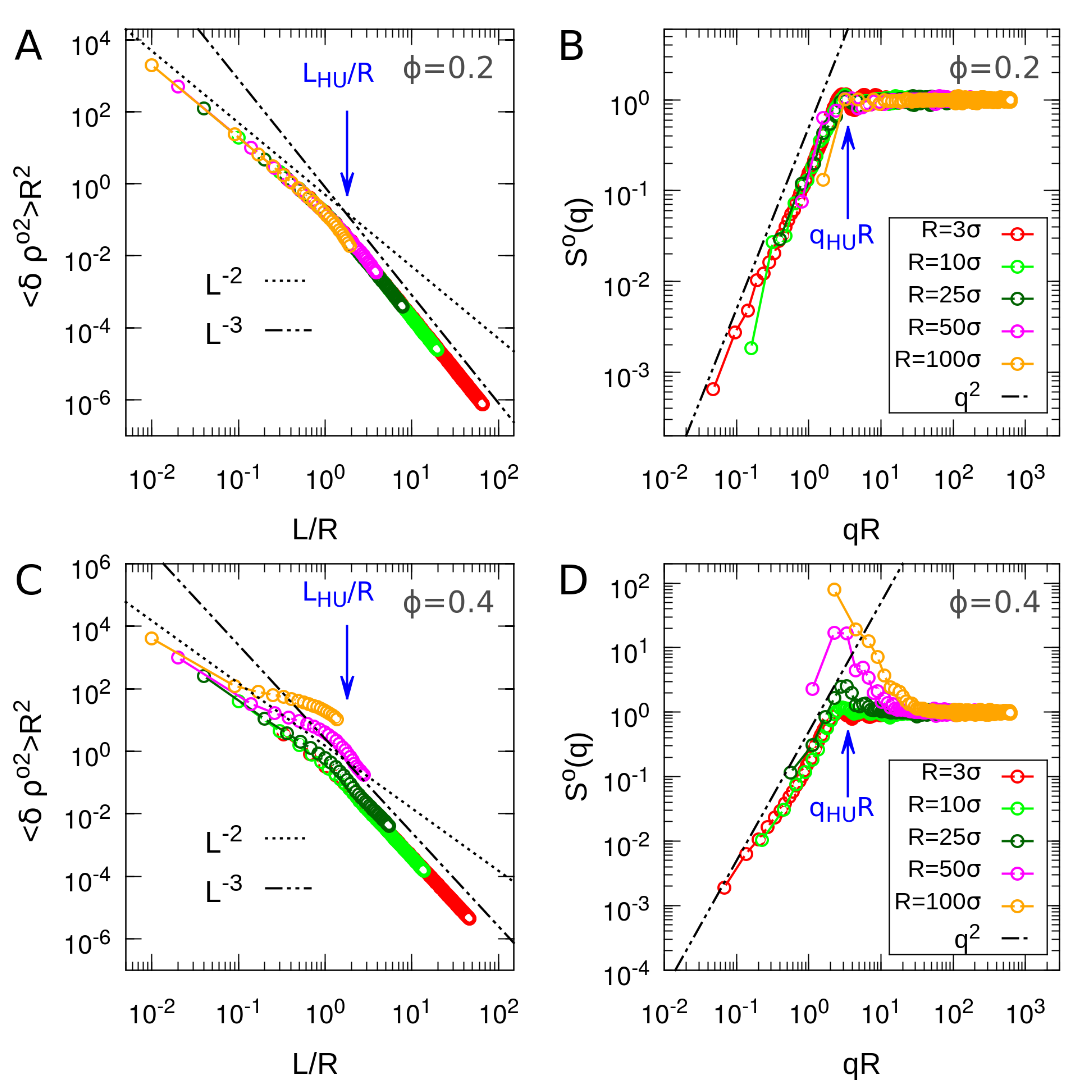} }
\end{tabular}
\caption{{\bf Global hyperuniformity.}  (\textbf{A,C}): Scaled density variance $\langle \delta \rho^{\mathrm o 2} \rangle R^2$ as functions of $L/R$, and (\textbf{B,D}): $S^{\mathrm o}(q)$ as  functions  of $qR$, for various $R$ at $\phi = 0.2$ (\textbf{A,B}) and $0.4$ (\textbf{C,D}), respectively.}
\label{Fig_collapse}
\end{figure*}

\subparagraph{Mechanism of global hyperuniformity}
To understand the hyperuniformity at large length scales, we focus on the spatial distribution of the instantaneous circling centers of active particles, which we treat as effective particles with radius $R$. For particle $i$ at position $\mathbf{r}_i$, the circling center is at $\mathbf{r}^{\mathrm o}_i =\mathbf{r}_i -\sigma^2 (\mathbf{F}^p_i \times \mathbf{\Omega})/|\mathbf{\Omega}|^2$, which is a fixed position for an isolated circle active particle. 
The motion of these effective particles is due to collisions with other particles. This is similar to what happens in the random organization model~\cite{hexner2015,tjhung2015,hexner2017noise}, where only overlapped particles experience random kicks. As indicated by the previous theoretical analysis, the self-trapping and the growing density fluctuations is confined within the length scale of $R$. At larger length scales, fluctuations decay as the diffusive mode (Fig.~\ref{Fig_dispersion}A). These imply that the dynamic equation for the density field of these effective particles $\rho^{\mathrm o}({\mathbf r},t)$ at large length scales can be described by the Fick's law of diffusion $\partial_t \rho = D \nabla^2 \rho$ in the $q$ space:
\begin{eqnarray}\label{Eq_cirlingCenter}
\partial_t \rho_{ \mathbf{q} }^{\mathrm o} = -D^{\mathrm o}_e q^2 \rho_{\mathbf{q} }^{\mathrm o} + \xi_{\mathbf{q}}(t)   ~~~~{\rm for}~~~~  \left(q \ll \frac{2\pi}{R} \right),
\end{eqnarray}
with $[\rho^{\mathrm o}_{\mathbf{q}},~\xi_{\mathbf{q}}] = \int d\mathbf{r} \cdot e^{-i\mathbf{q \cdot r}} [\rho^{\mathrm o},~\xi]$. Here, $D^{\mathrm o}_e$ is the diffusion coefficient of effective particles and $\xi_{\mathbf{q}}(t)$ is an additional noise term.
Since we neglect the thermal noise, $\xi_{\mathbf{q}}(t)$ comes from the chaotic multi-particle interaction~\cite{just2001}, which obeys the center of mass conservation (CMC). As proven in Ref~\cite{hexner2017noise}, the noise with additional CMC appears as a double space derivative in the diffusion equation, i.e., $\xi(t) = \sqrt{\overline{\rho}}~\nabla^2 \eta(t)$, with $\langle \eta(\mathbf{r},t)\eta(\mathbf{r}',t')\rangle = A^2  \delta(\mathbf{r} - \mathbf{r}')\delta(t-t')$ and $A$ the strength of the noise. This is different from the conventional single space derivative noise  that arises solely from the particle number conservation~\cite{dean1996langevin,marchetti2013}. Following Ref~\cite{hexner2017noise}, by adding such a double space derivative noise term into Eq.~\ref{Eq_cirlingCenter} as the perturbation, one obtains the dynamic equation for the density fluctuation $\delta \rho^{\mathrm o}$ in the Fourier space:
\begin{eqnarray}\label{Eq_cirlingCenter_noise}
i\omega \delta \rho_{ \mathbf{q}, \omega }^{\mathrm o} = -D^{\mathrm o}_e q^2 \delta \rho_{ \mathbf{q}, \omega }^{\mathrm o} - q^2 \sqrt{\overline{\rho}} ~\eta_{  \mathbf{q}, \omega } ~~~~~~   \left(q \ll \frac{2\pi}{R} \right),~~~~~~~~
\end{eqnarray}
where $\eta_{ \mathbf{q}, \omega }$ is the Fourier transform of the noise $\eta(\mathbf{r},t)$.
From Eq.~\ref{Eq_cirlingCenter_noise}, we obtain the structure factor
\begin{eqnarray}\label{Eq_cirlingCenter_Sk}
S^{\mathrm o}(\mathbf{q}) 
&=& \int^\infty_{-\infty} \frac{1}{2\pi \tau_{max}N}   \langle \delta \rho_{\mathbf{q},\omega}^{\mathrm o}  \delta \rho^{\mathrm o  *}_{\mathbf{q},\omega} \rangle d\omega \nonumber \\
&=& \frac{A^2}{2D_e}q^2~~~~~~~~~~~~~~  \left(q \ll \frac{2\pi}{R} \right),~~~~~~~~
\end{eqnarray}
where $\tau_{max}$ is the maximum observation time (see Supplementary Information S3). Eq.~\ref{Eq_cirlingCenter_Sk} shows the same hyperuniform exponent in $S(q)$ as observed in the simulations and also signatures the strongest hyperniform scaling for density variance $\langle \delta \rho^{\mathrm o 2} \rangle \sim L^{-3}$ at the length scales larger than $R$~\cite{torquato2003local}. In Fig.~\ref{Fig_collapse}A, we plot the scaled density variance $\langle \delta \rho^{\mathrm o 2} \rangle R^2$ versus $L/R$ for systems with various $R$ at $\phi = 0.2$, and one can see all points collapse into a single curve. The curve consists of two distinct scalings with a strongly hyperuniform scaling $L^{-3}$ at $L > L_{HU} \simeq 2R$ and a normal fluid-like scaling $L^{-2}$ at $L < L_{HU}$. Similar collapse is shown in Fig.~\ref{Fig_collapse}B for $S^{\mathrm o}(q)$ versus $qR$ where the hyperuniform scaling $S^{\mathrm o}(q)\sim q^2$ stops at $q_{HU} \simeq 2\pi/L_{HU}$.
For high density system, i.e., $\phi=0.4$, we obtain the same crossover length scale $L_{HU} \simeq 2R$ as shown in Fig.~\ref{Fig_collapse}C,D. In this case, the density variance and $S^{\mathrm o}(q)$ for different $R$ do not collapse into a single curve because of {cluster formation} induced large density fluctuations at $L \lesssim L_{HU}$. After obtaining the hyperuniform scaling for these effective particles, one can prove the existence of the same hyperuniform scaling at similar length scales for the circle active particles~\cite{gabrielli2002glass}.

\begin{figure*}[htbp] 
\centering
\begin{tabular}{c}
	\resizebox{120mm}{!}{\includegraphics[trim=0.0in 0.0in 0.0in 0.0in]{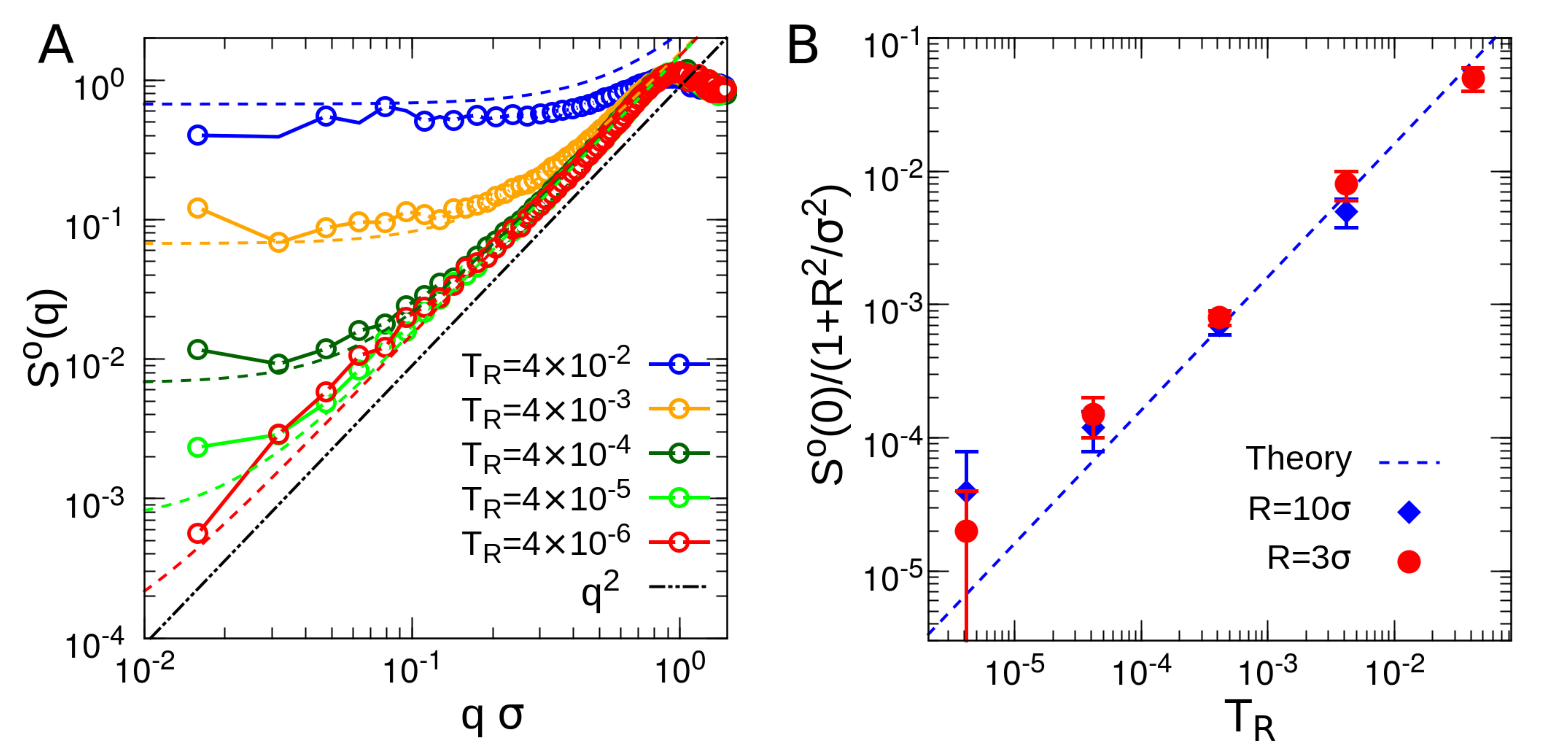} }
\end{tabular}
\caption{{\bf Effect of thermal noises on hyperuniformity.} (\textbf{A}) Structure factor $S^{\mathrm o}(q)$ under different reduced noise strength $T_{R}$ for $\phi=0.2$ and $R=3\sigma$. Open symbols show the simulation data, while the dashed line is the theoretical prediction of Eq.\ref{eq_thermal}. (\textbf{B})~Normalized $S^{\mathrm o}(0)$ as functions of $T_{R}$ from theoretical prediction (dashed line) and the fitting of simulation results (solid symbols) for systems with $\phi=0.2$. For all the calculations, $N = 40,000$.}
\label{Fig_noise}
\end{figure*}

From these analyses, one can see that in systems of circle active particles, there are two ingredients for the global hyperuniformity: i) the circular trajectories of active particles localize their active motions, creating the {Fickian} diffusion condition for the active particles at large length scales; ii) the local density fluctuations originate from the chaotic multi-particle interactions which obey the CMC. {These arguments are  based on Ref~\cite{hexner2017noise}, which produces the same hyperuniform scaling in $S(q)$ using the random-organization model with CMC. Although the two systems share the similar mechanism of hyperuniformity at large length scales, there are also marked differences between them. First, the driving force in our system is persistent and the dynamics of the active particles is overdamped and deterministic. Second, our system has a characteristic length $R$, which separates two completely different fluctuations. These can not be described by the model in Ref~\cite{hexner2017noise}.}

We notice the  two ingredients for the dynamic hyperuniformity above seem to be satisfied in a recent experimental system, in which spherical Janus colloids with additional soft repulsion perform almost perfect circling motion with the same chirality but two different phases, driving by external fields~\cite{han2017effective}. The thermal noise was believed to be negligible in this system and the interactions between these nearly athermal active particles produce an effective temperature which controls both the kinetic and phase behavior of the system~\cite{han2017effective}. To test whether the hyperuniformity can exist in this circle active particle system with the bimodal circling-phase distribution, we simulate a low density system ($\phi=0.2$) with two different circle radii $R=1\sigma,~ 3\sigma$ using the model (See Method) which produced almost the same experimental result in Ref.~\cite{han2017effective}. Our result is shown in Supplementary Fig.~S4, from which we find the bimodal-distributed system falls into a phase-separated absorbing state for $R=1\sigma$, but stays in an active mixing (lane) state for $R=3 \sigma$, consistent with the previous finding~\cite{han2017effective}. In the active mixing state, we observe the same hyperuniform scaling $S(q\rightarrow 0)\sim q^2$. This result demonstrates the robustness of the hyperuniformity mechanism unveiled by this general model and suggests a high possibility of realization in experiments.  Moreover, it also suggests that other active systems satisfying the two ingredients, like active spinner systems~\cite{scholz2017velocity} or size/interaction oscillating particle systems~\cite{tjhung2017}, may be used to produce the same hyperuniformity as well.

\subparagraph{Effect of thermal noises}
According to the definition, perfectly hyperuniform state requires $S(0)$ to exactly equal zero. However, in real experimental systems, thermal noise is unavoidable. Based on the fluctuation-compressibility relationship
\begin{eqnarray}\label{eq_S0}
S(0)=\kappa_T \rho k_BT,
\end{eqnarray}
any thermal equilibrated systems with the positive isothermal compressibility $\kappa_T$ at finite temperatures can not be strictly hyperuniform due to thermal excitation~\cite{kim2018effect}. In crystals, thermal excitation appears as phonon modes, which cause background scattering or thermal diffuse scattering, whose effect can be measured by Debye-Waller factor~\cite{kim2018effect,rundquist1991thermal}. Nevertheless, the wide application of crystal materials indicates that thermalization only weakens but does not destroy most physical properties of the ground-state crystal. Similarly, \emph{nearly} hyperuniformity~\cite{torquato2018PR,kim2018effect,xie2013silicon} in disordered structures may also be enough to achieve some desired functions, e.g., isotropic photonic/electronic band gaps~\cite{edagawa2008photonic,xie2013silicon}. In Fig.~\ref{Fig_noise}, we analyze the influence of thermal excitation on the ground-state hyperuniform system with $\phi=0.2$ and $R=3\sigma$. We find with gradual increase of the reduced noise strength $T_{R}$ from zero, $S(q\rightarrow 0)$ begins to saturate at some non-zero value which increases along with $T_{R}$ (Fig.~\ref{Fig_noise}A). In Supplementary Information S4, we introduce the thermal noise ${\mathbf f}=[f_x,f_y]$ in Eq. \ref{Eq_cirlingCenter} as,
\begin{eqnarray} 
\xi(t) = \sqrt{\overline{\rho}} ~\nabla \cdot~[\nabla \eta(t) +  {\mathbf f}(t) ],
\end{eqnarray}
where $\langle f_i(\mathbf{r},t)f_j(\mathbf{r}',t')\rangle =2  D_{therm}\delta_{ij}\delta(\mathbf{r} - \mathbf{r}')\delta(t-t') $ with\ $D_{therm}=k_BT \gamma_t^{-1}\left( 1 + {R^2}/{\sigma^2}\right)$ the self-diffusion constant of effective particles due to the thermal Brownian motion~\cite{lowen2016}. Here we assume that the  thermal noise is a first-order weak perturbation on the chaotic noise $\eta(t)$, which leads to the decoupling of these two noise sources: $\langle f_i(\mathbf{r},t)\eta(\mathbf{r'},t')\rangle =0$. Then we can estimate $S^{\mathrm o}(q)$ as a function of the reduced noise strength $T_{R}$ for low density systems as
\begin{eqnarray} \label{eq_thermal}
S^{\mathrm o}({q}) &=& \frac{v_0 \sigma} { D_e}\left( 1 + \frac{R^2}{\sigma^2}\right) T_{R} + \frac{A^2  }{2 D_e  } q^2~~~ \left(q < q_{HU} \right).~~~~~~~~
\end{eqnarray}
In Fig.~\ref{Fig_noise}A, we plot the theoretical prediction of Eq.~\ref{eq_thermal}  as dashed lines for different $T_{R}$ at $R=3\sigma$ and $\phi=0.2$, by assuming them all across the same $S^{\mathrm o}(q_{HU})$ point. In Fig.~\ref{Fig_noise}B, we compare the $S^{\mathrm o}(0)$ from the theoretical prediction with the  $S^{\mathrm o}(0)$ obtained from the fitting of simulation results (see Method) for systems with $R=3\sigma$ and $R=10\sigma$ at $\phi=0.2$. 
One can find quantitative agreements between the simulation and theoretical predictions in Fig.~\ref{Fig_noise}. which suggest that the susceptibility of hyperuniformity to the noise in our non-equilibrium system is similar to that in thermalized crystals, i.e., Eq.~\ref{eq_S0}. { However, we also emphasize the difference: in our non-equilibrium system, the saturated value of $S^{\mathrm o}(0)$ is determined by the driving force as well. Therefore, in experiments it is possible to observe a large range of hyperuniform scaling in $S(q)$ or density variance as long as the driving force is much larger than the thermal noise, i.e., $T_R\ll 1$.}


\section{Conclusions}
In conclusion, by combining computer simulations with theoretical analyses, we investigate the dynamic phase behaviors in 2D systems of circle active particles. {In the zero-noise limit}, we find that with increasing the density of system or the radius of circular motion $R$, the system undergoes a transition from an absorbing state to an active fluid state, which is accompanied with a structural transformation {for small $R$}. In the active fluid state, we find a characteristic length scale $L_{HU}$. For $L \gg L_{HU}$, the system exhibits strong hyperuniformity with the density variance scaling as the same as in perfect crystals, while for $L \lesssim L_{HU}$, we observe normal random fluctuations at  low  density and large density fluctuations {(cluster formation)} at  high density. To understand the mechanism of the phase behaviors of the system, we develop a dynamic mean-field theory. Linear stability analysis suggests that {at the mean-field level,} the large local density fluctuations at relative large $R$ are a result of motility-induced micro-phase separation which is confined within the length scale of $R$. {For the global hyperuniformity, we attribute it to} the interplay between the {Fickian} diffusion of active particles at large length scales and local particle collisions that conserve the center of mass{~\cite{hexner2017noise}}. Our work demonstrates that two extreme fluctuations, i.e., {large density fluctuations} and hyperuniformity can coexist in the same dynamic system at different length scales.  { We emphasize that this stable hierarchical  hyerunifrom fluid is conceptually different from the disordered hyperuniform solid or critical hyperuniform state.}  From a practical point of view, our results suggest that even for exotic disordered hyperuniform structures, there is still plenty of room at the `bottom', i.e., one may construct arbitrary local complex structures (ordered or disordered) with extra functionalities without harming the global hyperuniformity. This provides large freedom in designing hierarchical disordered hyperuniform materials with unconventional properties.


\section{Methods}
\noindent
In our simulations, we use a square simulation box with periodic boundary conditions in all directions, starting from initial configurations with random particle positions and orientations {to make sure the initial structure factor $S(q)\sim 1$}. The time unit is chosen to be the time that a particle moves a distance of $\sigma$ in the dilute limit, i.e.,  $\tau_0=\sigma/v_0$. To mimic the excluded volume interaction between colloidal particles $i$ and $j$, we employ the Weeks-Chandler-Andersen (WCA) potential
\begin{eqnarray}
{U(r_{ij})}=\left\{
\begin{array}{lr}
4\epsilon \left[\left(\frac{\sigma}{r_{ij}}\right)^{12}-\left(\frac{\sigma}{r_{ij}}\right)^{6}+\frac{1}{4}\right] & (r_{ij}<2^{1/6}\sigma)~~~~~~~\\
0& (r_{ij}>2^{1/6}\sigma)~~~~~~~
\end{array}
\right.
\end{eqnarray}
where  $r_{ij}$ is the center to center distance between particle $i$ and $j$ with $\sigma$ the diameter of particles. We choose $\epsilon=F^p\sigma/24$, which gives the typical contact distance $ r_c = \sigma$ between particles based on force balance $F^p = \left.\frac{\partial U(r_{ij})}{\partial r_{ij}}\right|_{r_{ij}=r_c}$.  For systems at $T_{R}=0$, to exclude the noise generated by discrete dynamic integrations,  we use perfect convex polygons to approximate the closed circle trajectories of active particles.  This is realized by finely tuning the integration step which is around $10^{-3} \tau_0$. 

For systems with bimodal distributed circling-phase, following Ref.~\cite{han2017effective}, we add an additional soft repulsion $U_s = A~ \left({r}/{\sigma}\right)^{-4}$ with cutoff distance $r_{sc}=5\sigma$ and $A=7.5 \epsilon$ to model the isotropic dipole interaction between active particles. The self-propulsion force for this system is reset to $F^p = {2^{7/6} A}/{\sigma}$ to make the dipolar interaction balance the driving force at $r_c=2^{1/6}\sigma$~\cite{han2017effective}. 

The density variance $\langle \delta \rho^2 \rangle$ of the system is calculated using a spherical window whose diameter is smaller than the half of simulation box to avoid the finite size effect. Under the periodic boundary condition, the structure factor $S(q)$ is calculated for some discrete $\mathbf q$ vectors, i.e., $[q_x,q_y]= \frac{2\pi}{L_0} [i ,j]~(i,j=1,2,3 \cdots)$ with $L_0$ the size of cubic simulation box. {The simulation time for the equilibrating and sampling processes depend on the system size and density. For hyperuniform system, the criteria for the system to reach equilibrium is whether the hyperuniform scaling of $S(q)$ has fully extended to the smallest $q_{min}=2\pi/L_0$ without further change.} The fitting function used in Fig.~\ref{Fig_noise}B is Eq.~\ref{eq_thermal} with adjustable $T_{R}$.

\section{Supplementary Materials}
\noindent
Supplementary material for this article is available at XXXXX-XXXXX 
\\

\begin{small}
\noindent
Section S1. Derivation of  the dynamic mean-field theory for 2D system of circle active particles.\\

\noindent
Section S2. Linear stability analysis.\\

\noindent
Section S3. Calculation of $ S^{\mathrm o}(q)$ for system with CMC.\\

\noindent
Section S4. Effect of thermal noise on  $ S^o(q)$.\\

\noindent
{Figure S1. Size independence of exponent $\beta$ for diffusion coefficient $D$ at $R=1.75\sigma$.}\\

\noindent
{Figure S2. Structural comparison between active state and absorbing state near the critical point at $R=10\sigma$.}\\

\noindent
{Figure S3. Structure factor and typical configurations for large systems with $N=102,400$ at $\phi=0.4$ under different $R$.}\\

\noindent
Figure S4. Hyperunifomity in an experimentally realizable system with bimodal circling-phase distribution.\\

\noindent
Movie 1-2. Absorbing state and active state  in Fig.~\ref{Fig_schematic}C,D.\\

\noindent
Movie 3-7. Active states with different $R$ in Fig.~\ref{Fig_hyperuniform}G.
\end{small}


\begin{thebibliography}{0}%
\makeatletter
\providecommand \@ifxundefined [1]{%
 \@ifx{#1\undefined}
}%
\providecommand \@ifnum [1]{%
 \ifnum #1\expandafter \@firstoftwo
 \else \expandafter \@secondoftwo
 \fi
}%
\providecommand \@ifx [1]{%
 \ifx #1\expandafter \@firstoftwo
 \else \expandafter \@secondoftwo
 \fi
}%
\providecommand \natexlab [1]{#1}%
\providecommand \enquote  [1]{``#1''}%
\providecommand \bibnamefont  [1]{#1}%
\providecommand \bibfnamefont [1]{#1}%
\providecommand \citenamefont [1]{#1}%
\providecommand \href@noop [0]{\@secondoftwo}%
\providecommand \href [0]{\begingroup \@sanitize@url \@href}%
\providecommand \@href[1]{\@@startlink{#1}\@@href}%
\providecommand \@@href[1]{\endgroup#1\@@endlink}%
\providecommand \@sanitize@url [0]{\catcode `\\12\catcode `\$12\catcode
  `\&12\catcode `\#12\catcode `\^12\catcode `\_12\catcode `\%12\relax}%
\providecommand \@@startlink[1]{}%
\providecommand \@@endlink[0]{}%
\providecommand \url  [0]{\begingroup\@sanitize@url \@url }%
\providecommand \@url [1]{\endgroup\@href {#1}{\urlprefix }}%
\providecommand \urlprefix  [0]{URL }%
\providecommand \Eprint [0]{\href }%
\providecommand \doibase [0]{http://dx.doi.org/}%
\providecommand \selectlanguage [0]{\@gobble}%
\providecommand \bibinfo  [0]{\@secondoftwo}%
\providecommand \bibfield  [0]{\@secondoftwo}%
\providecommand \translation [1]{[#1]}%
\providecommand \BibitemOpen [0]{}%
\providecommand \bibitemStop [0]{}%
\providecommand \bibitemNoStop [0]{.\EOS\space}%
\providecommand \EOS [0]{\spacefactor3000\relax}%
\providecommand \BibitemShut  [1]{\csname bibitem#1\endcsname}%
\let\auto@bib@innerbib\@empty
\end{thebibliography}%


\begin{thebibliography}{40}
\expandafter\ifx\csname natexlab\endcsname\relax\def\natexlab#1{#1}\fi
\expandafter\ifx\csname url\endcsname\relax
  \def\url#1{\texttt{#1}}\fi
\expandafter\ifx\csname urlprefix\endcsname\relax\def\urlprefix{URL }\fi

\bibitem[{Torquato(2018)}]{torquato2018PR}
Torquato, S.
\newblock Hyperuniform states of matter.
\newblock \emph{Phys. Rep.}  (2018).

\bibitem[{Torquato \& Stillinger(2003)}]{torquato2003local}
Torquato, S. \& Stillinger, F.~H.
\newblock Local density fluctuations, hyperuniformity, and order metrics.
\newblock \emph{Phys. Rev. E} \textbf{68}, 041113 (2003).

\bibitem[{Donev \emph{et~al.}(2005)Donev, Stillinger \&
  Torquato}]{donevprl2005}
Donev, A., Stillinger, F.~H. \& Torquato, S.
\newblock Unexpected density fluctuations in jammed disordered sphere packings.
\newblock \emph{Phys. Rev. Lett.} \textbf{95}, 090604 (2005).

\bibitem[{Jiao \emph{et~al.}(2014)}]{jiaopre2014}
Jiao, Y. \emph{et~al.}
\newblock Avian photoreceptor patterns represent a disordered hyperuniform
  solution to a multiscale packing problem.
\newblock \emph{Phys. Rev. E} \textbf{89}, 022721 (2014).

\bibitem[{Hexner \& Levine(2015)}]{hexner2015}
Hexner, D. \& Levine, D.
\newblock Hyperuniformity of critical absorbing states.
\newblock \emph{Phys. Rev. Lett.} \textbf{114}, 110602 (2015).

\bibitem[{Tjhung \& Berthier(2015)}]{tjhung2015}
Tjhung, E. \& Berthier, L.
\newblock Hyperuniform density fluctuations and diverging dynamic correlations
  in periodically driven colloidal suspensions.
\newblock \emph{Phys. Rev. Lett.} \textbf{114}, 148301 (2015).

\bibitem[{Hexner \emph{et~al.}(2017)Hexner, Chaikin \&
  Levine}]{hexner2017enhanced}
Hexner, D., Chaikin, P.~M. \& Levine, D.
\newblock Enhanced hyperuniformity from random reorganization.
\newblock \emph{Proc. Natl. Acad. Sci. U.S.A} \textbf{114}, 4294 (2017).

\bibitem[{Weijs \emph{et~al.}(2015)Weijs, Jeanneret, Dreyfus \&
  Bartolo}]{weijs2015emergent}
Weijs, J.~H., Jeanneret, R., Dreyfus, R. \& Bartolo, D.
\newblock Emergent hyperuniformity in periodically driven emulsions.
\newblock \emph{Phys Rev Lett} \textbf{115}, 108301 (2015).

\bibitem[{Wang \emph{et~al.}(2018)Wang, Schwarz \&
  Paulsen}]{wang2017hyperuniformity}
Wang, J., Schwarz, J. \& Paulsen, J.~D.
\newblock Hyperuniformity with no fine tuning in sheared sedimenting
  suspensions.
\newblock \emph{Nat. Commun.} \textbf{9}, 2836 (2018).

\bibitem[{Hexner \& Levine(2017)}]{hexner2017noise}
Hexner, D. \& Levine, D.
\newblock Noise, diffusion, and hyperuniformity.
\newblock \emph{Phys. Rev. Lett.} \textbf{118}, 020601 (2017).

\bibitem[{Florescu \emph{et~al.}(2009)Florescu, Torquato \&
  Steinhardt}]{florescu2009designer}
Florescu, M., Torquato, S. \& Steinhardt, P.~J.
\newblock Designer disordered materials with large, complete photonic band
  gaps.
\newblock \emph{Proc. Natl. Acad. Sci. U.S.A} \textbf{106}, 20658--20663
  (2009).

\bibitem[{Man \emph{et~al.}(2013)}]{man2013photonic}
Man, W. \emph{et~al.}
\newblock Photonic band gap in isotropic hyperuniform disordered solids with
  low dielectric contrast.
\newblock \emph{Opt. Express} \textbf{21}, 19972--19981 (2013).

\bibitem[{Kim \& Torquato(2018)}]{kim2018effect}
Kim, J. \& Torquato, S.
\newblock Effect of imperfections on the hyperuniformity of many-body systems.
\newblock \emph{Phys. Rev. B} \textbf{97}, 054105 (2018).

\bibitem[{Edagawa \emph{et~al.}(2008)Edagawa, Kanoko \&
  Notomi}]{edagawa2008photonic}
Edagawa, K., Kanoko, S. \& Notomi, M.
\newblock Photonic amorphous diamond structure with a 3D photonic band gap.
\newblock \emph{Phys. Rev. Lett.} \textbf{100}, 013901 (2008).

\bibitem[{Xie \emph{et~al.}(2013)}]{xie2013silicon}
Xie, R. \emph{et~al.}
\newblock Hyperuniformity in amorphous silicon based on the measurement of the
  infinite-wavelength limit of the structure factor.
\newblock \emph{Proc. Natl. Acad. Sci. U.S.A} \textbf{110}, 13250--13254
  (2013).

\bibitem[{Leseur \emph{et~al.}(2016)Leseur, Pierrat \&
  Carminati}]{leseur2016high}
Leseur, O., Pierrat, R. \& Carminati, R.
\newblock High-density hyperuniform materials can be transparent.
\newblock \emph{Optica} \textbf{3}, 763--767 (2016).

\bibitem[{Corte \emph{et~al.}(2008)Corte, Chaikin, Gollub \&
  Pine}]{corte2008random}
Corte, L., Chaikin, P., Gollub, J.~P. \& Pine, D.
\newblock Random organization in periodically driven systems.
\newblock \emph{Nat. Phys.} \textbf{4}, 420--424 (2008).

\bibitem[{Marchetti \emph{et~al.}(2013)}]{marchetti2013}
Marchetti, M.~C. \emph{et~al.}
\newblock Hydrodynamics of soft active matter.
\newblock \emph{Rev. Mod. Phys.} \textbf{85}, 1143 (2013).

\bibitem[{Ramaswamy \emph{et~al.}(2003)Ramaswamy, Simha \&
  Toner}]{ramaswamy2003active}
Ramaswamy, S., Simha, R.~A. \& Toner, J.
\newblock Active nematics on a substrate: Giant number fluctuations and
  long-time tails.
\newblock \emph{EPL (Europhysics Letters)} \textbf{62}, 196 (2003).

\bibitem[{Fily \& Marchetti(2012)}]{fily2012athermal}
Fily, Y. \& Marchetti, M.~C.
\newblock Athermal phase separation of self-propelled particles with no
  alignment.
\newblock \emph{Phys. Rev. Lett.} \textbf{108}, 235702 (2012).

\bibitem[{Speck \emph{et~al.}(2014)Speck, Bialk{\'e}, Menzel \&
  L{\"o}wen}]{speck2014effective}
Speck, T., Bialk{\'e}, J., Menzel, A.~M. \& L{\"o}wen, H.
\newblock Effective Cahn-Hilliard equation for the phase separation of active
  Brownian particles.
\newblock \emph{Phys. Rev. Lett.} \textbf{112}, 218304 (2014).

\bibitem[{Ni \emph{et~al.}(2015)Ni, Stuart \& Bolhuis}]{niprl2015}
Ni, R., Stuart, M. A.~C. \& Bolhuis, P.~G.
\newblock Tunable long range forces mediated by self-propelled colloidal hard
  spheres.
\newblock \emph{Phys. Rev. Lett.} \textbf{114}, 018302 (2015).

\bibitem[{Buttinoni \emph{et~al.}(2013)}]{lowen2013}
Buttinoni, I. \emph{et~al.}
\newblock Dynamical clustering and phase separation in suspensions of
  self-propelled colloidal particles.
\newblock \emph{Phys. Rev. Lett.} \textbf{110}, 238301 (2013).

\bibitem[{Lauga \emph{et~al.}(2006)Lauga, DiLuzio, Whitesides \&
  Stone}]{lauga2006}
Lauga, E., DiLuzio, W.~R., Whitesides, G.~M. \& Stone, H.~A.
\newblock Swimming in circles: motion of bacteria near solid boundaries.
\newblock \emph{Biophys. J.} \textbf{90}, 400--412 (2006).

\bibitem[{L{\"o}wen(2016)}]{lowen2016}
L{\"o}wen, H.
\newblock Chirality in microswimmer motion: From circle swimmers to active
  turbulence.
\newblock \emph{The European Physical Journal Special Topics} \textbf{225},
  2319--2331 (2016).

\bibitem[{Han \emph{et~al.}(2017)Han, Yan, Granick \&
  Luijten}]{han2017effective}
Han, M., Yan, J., Granick, S. \& Luijten, E.
\newblock Effective temperature concept evaluated in an active colloid mixture.
\newblock \emph{Proc. Natl. Acad. Sci. U.S.A} \textbf{114}, 7513--7518 (2017).

\bibitem[{Zhou \emph{et~al.}(2017)Zhou, Zhao, Wei \& Wang}]{zhou2017twists}
Zhou, C., Zhao, L., Wei, M. \& Wang, W.
\newblock Twists and Turns of Orbiting and Spinning Metallic Microparticles
  Powered by Megahertz Ultrasound.
\newblock \emph{ACS Nano} \textbf{11}, 12668--12676 (2017).

\bibitem[{Wioland \emph{et~al.}(2016)Wioland, Woodhouse, Dunkel \&
  Goldstein}]{wioland2016}
Wioland, H., Woodhouse, F.~G., Dunkel, J. \& Goldstein, R.~E.
\newblock Ferromagnetic and antiferromagnetic order in bacterial vortex
  lattices.
\newblock \emph{Nat. Phys.} \textbf{12}, 341 (2016).

\bibitem[{Scholz \& P{\"o}schel(2017)}]{scholz2017velocity}
Scholz, C. \& P{\"o}schel, T.
\newblock Velocity distribution of a homogeneously driven two-dimensional
  granular gas.
\newblock \emph{Phys. Rev. Lett.} \textbf{118}, 198003 (2017).

\bibitem[{Ma \emph{et~al.}(2017)Ma, Lei \& Ni}]{ma2017driving}
Ma, Z., Lei, Q.-l. \& Ni, R.
\newblock Driving dynamic colloidal assembly using eccentric self-propelled
  colloids.
\newblock \emph{Soft matter} \textbf{13}, 8940--8946 (2017).

\bibitem[{Liebchen \& Levis(2017)}]{liebchen2017collective}
Liebchen, B. \& Levis, D.
\newblock Collective behavior of chiral active matter: pattern formation and
  enhanced flocking.
\newblock \emph{Phys. Rev. Lett.} \textbf{119}, 058002 (2017).

\bibitem[{Chen \emph{et~al.}(2017)Chen, Liu, Shi, Chat{\'e} \&
  Wu}]{chen2017weak}
Chen, C., Liu, S., Shi, X.-q., Chat{\'e}, H. \& Wu, Y.
\newblock Weak synchronization and large-scale collective oscillation in dense
  bacterial suspensions.
\newblock \emph{Nature} \textbf{542}, 210 (2017).

\bibitem[{Souslov \emph{et~al.}(2017)Souslov, van Zuiden, Bartolo \&
  Vitelli}]{souslov2017topological}
Souslov, A., van Zuiden, B.~C., Bartolo, D. \& Vitelli, V.
\newblock Topological sound in active-liquid metamaterials.
\newblock \emph{Nat. Phys.} \textbf{13}, 1091 (2017).

\bibitem[{Hinrichsen(2000)}]{hinrichsen2000non}
Hinrichsen, H.
\newblock Non-equilibrium critical phenomena and phase transitions into
  absorbing states.
\newblock \emph{Adv. Phys.} \textbf{49}, 815--958 (2000).

\bibitem[{L{\"u}beck(2003)}]{lubeck2003}
L{\"u}beck, S.
\newblock Universal behavior of crossover scaling functions for continuous
  phase transitions.
\newblock \emph{Phys. Rev. Lett.} \textbf{90}, 210601 (2003).

\bibitem[{Just \emph{et~al.}(2001)Just, Kantz, R{\"o}denbeck \&
  Helm}]{just2001}
Just, W., Kantz, H., R{\"o}denbeck, C. \& Helm, M.
\newblock Stochastic modelling: replacing fast degrees of freedom by noise.
\newblock \emph{J. Phys. A: Math. Gen.} \textbf{34}, 3199 (2001).

\bibitem[{Dean(1996)}]{dean1996langevin}
Dean, D.~S.
\newblock Langevin equation for the density of a system of interacting Langevin
  processes.
\newblock \emph{Journal of Physics A: Mathematical and General} \textbf{29},
  L613 (1996).

\bibitem[{Gabrielli \emph{et~al.}(2002)Gabrielli, Joyce \&
  Labini}]{gabrielli2002glass}
Gabrielli, A., Joyce, M. \& Labini, F.~S.
\newblock Glass-like universe: Real-space correlation properties of standard
  cosmological models.
\newblock \emph{Phys. Rev. D} \textbf{65}, 083523 (2002).

\bibitem[{Tjhung \& Berthier(2017)}]{tjhung2017}
Tjhung, E. \& Berthier, L.
\newblock Discontinuous fluidization transition in time-correlated assemblies
  of actively deforming particles.
\newblock \emph{Phys. Rev. E} \textbf{96}, 050601 (2017).

\bibitem[{Rundquist \emph{et~al.}(1991)Rundquist, Kesavamoorthy, Jagannathan \&
  Asher}]{rundquist1991thermal}
Rundquist, P.~A., Kesavamoorthy, R., Jagannathan, S. \& Asher, S.~A.
\newblock Thermal diffuse scattering from colloidal crystals.
\newblock \emph{J. Chem. Phys.} \textbf{95}, 1249--1257 (1991).

\end{thebibliography}

\ \\
\begin{acknowledgments}
\textbf{Acknowledgments:} 
We thank Prof. Hepeng Zhang in Shanghai Jiao Tong University for fruitful discussions. We are grateful to the National Supercomputing Centre (NSCC) of Singapore for supporting the numerical calculations. \textbf{Funding:} This work is supported by Nanyang Technological University Start-Up Grant (NTU-SUG: M4081781.120), the Academic Research Fund from Singapore Ministry of Education (M4011616.120, M4011873.120, and MOE2017-T2-1-066 (S)), and the Advanced Manufacturing and Engineering Young Individual Research Grant (A1784C0018) by the Science and Engineering Research Council of Agency for Science, Technology and Research Singapore. \textbf{Author contributions:} Q.-L.L. and R.N. conceived the research; Q.-L.L. performed the research; R.N. directed the research; Q.-L.L, M.P.C. and R.N. analysed the data and wrote the manuscript. \textbf{Competing interests:} The authors declare that they have no competing interests. \textbf{Data and materials availability:} All data needed to evaluate the conclusions in the paper are presented in the paper and/or the Supplementary Materials. Additional data related to this paper may be requested from the authors.
\end{acknowledgments}

\end{document}